\documentclass[12pt,preprint]{aastex}
\shortauthors{Hawley \& Johns--Krull} 
\shorttitle{Transition Region Emission from VLM Stars}
\begin{document}

\title{Transition Region Emission From Very Low Mass Stars}

\author{Suzanne L. Hawley\altaffilmark{1},
Christopher M. Johns--Krull\altaffilmark{2}}
\altaffiltext{1}{Astronomy Department, University of Washington,
   Box 351580, Seattle, WA  98195\\
email: slh@astro.washington.edu}
\altaffiltext{2}{Department of Physics \& Astronomy, Rice University, 
   6100 Main St., Houston, TX  77005\\
email: cmj@rice.edu}


\begin{abstract}

We present results from our cycle 10 HST program to
search for transition region emission in very low-mass,
main-sequence
stars in the spectral range M7-M9.  Our program is aimed
at 1) detecting emission; and 2) distinguishing between
flaring and quiescent origin for the emission.  We have
obtained HST/STIS time series observations of three
active, very low mass stars (VB 8, VB 10, and LHS 2065) which
show persistent activity in transition region lines including
\ion{Si}{4}, \ion{C}{4}, and \ion{He}{2}.  Emission in transition
region lines appears to be variable between exposures, but
is always observed.  A strong flare was also observed in one 
10 minute exposure on VB 10.  Our results indicate
that active, very low-mass stars maintain a persistent quiescent
chromosphere and transition region that is similar to those observed
in active, earlier type M dwarfs, in contrast to suggestions
that these low-mass, main-sequence stars exhibit only relatively
strong flares and no quiescent emission.

\end{abstract}



\section{Introduction}

Surface magnetic fields on low mass stars give rise to a
variety of interesting astrophysical phenomena, including
a persistent hot outer atmosphere (which may be contained
within magnetic loop structures) and transient
events such as starspots and flares.  Non-radiative heating
processes transfer energy from the magnetic field into
the outer stellar atmosphere to power these phenomena, but
a physical description of the heating processes remains
elusive even on the Sun.  Observationally, the heating
produces chromospheric (T$\sim 10^4$K), 
transition region (T$\sim 10^5$K)
and coronal (T$\sim 10^6$K) emission, 
primarily in emission lines of
the hydrogen Lyman and Balmer series, and in ionized
species of abundant elements such as helium, carbon,
oxygen, calcium, magnesium, and iron.  

Recently, there has been renewed interest in the atmospheric
properties of the lowest mass stars and brown dwarfs.  Chromospheric
observations of H$\alpha$ emission \citep{h96,g00,bu02} indicate
that both the frequency of active stars at a given spectral
type, and the activity level (measured by L$_{H\alpha}/L_{bol}$)
decrease at spectral types later than approximately M7.  Recent
investigations of magnetic activity in these very late type dwarfs
have led to a new, yet already widely accepted, paradigm:
that hot gas in the atmospheres of 
very low mass main sequence stars and brown dwarfs
is present only during flares \citep{f00,r00,mb02}.
Here, hot gas refers to gas at transition region
and coronal temperatures.  The M9 brown dwarf, LP944-20, observed
with Chandra by
\citet{r00} does show persistent chromospheric (H$\alpha$) 
emission during quiescence (i.e. outside of flares), as do 
many of its counterparts.

This new paradigm is based on the following evidence.
\citet{l95} observed the M8 dwarf VB 10 with HST/GHRS and
found no quiescent transition region emission, but did observe
\ion{C}{4} emission during a flare.  \citet{f00} re-analyzed
ROSAT data on VB10 and found that the observed X-ray emission
came only from a flare event, and no X-rays were detected during
six hours of quiescent observation.  The X-ray flare luminosity
was substantially larger than expected based on the upper limit
of the quiescent X-ray luminosity, when compared to flares on
earlier type (M0 - M6) stars.  An H$\alpha$ flare on an M9.5 2MASS dwarf
showed a similarly substantial flare luminosity compared to
quiescence \citep{l99}.  Finally, the afore-mentioned
Chandra investigation of the M9 brown dwarf LP944-20 detected no X-rays
in a quiescent observation of nine hour duration \citep{r00}.  By contrast,
\citet{b01} found both persistent and flaring radio emission
from LP944-20, while \citet{b02} recently reported radio 
observations of both flaring and quiescent emission 
for 3 additional low mass dwarfs of spectral types M8.5, M9.5 and L3.5.  
These authors speculate that perhaps the radio emission mechanism
differs in the low mass dwarfs compared to earlier type active stars.
On the theoretical side, both \citet{f00} and \citet{mb02} have
proposed mechanisms related to the low ionization state of very cool 
atmospheres which would prevent the occurrence of magnetic
activity in very cool (hence late spectral type) dwarfs, in order 
to explain the various observations described above.

Recent X-ray studies have probed into the brown dwarf regime in
young clusters \citep{c00,f02} and
find both quiescent and flaring emission.  However, the extreme youth of
these objects means that most have early-mid M spectral types.  Further,
fossil magnetic fields may play a significant role in their activity
\citep{f02}, making them unsuitable as a testbed for
theories of the evolution of dynamo-generated magnetic fields
in older, main sequence dwarfs.  

Extensive observations of earlier M dwarfs (type M0-M6) have
led to scaling relations between the persistent H$\alpha$ 
(chromospheric) and soft X-ray (thermal coronal) emission outside of
flares \citep{h96}.  We have used H$\alpha$ observations \citep{h96} and
\ion{C}{4} (transition region) measurements \citep{j00} for the dMe stars
AD Leo (M3), EV Lac (M3.5), and YZ CMi (M4) to derive empirical \ion{C}{4}
to H$\alpha$ line flux ratios of 0.06, 0.07, and 0.13 respectively.
There may be a tendency for this ratio to increase toward later
spectral types.  A conservative estimate suggests that the ratio is
typically $\sim$ 0.1.  In terms of the observed
fluxes, the relations between chromospheric, transition region, and
coronal diagnostics for active early-mid M dwarfs are therefore:
$$  F_{C IV } \sim  0.1 F_{H\alpha} \eqno(1)$$
$$  F_X \sim  3 F_{H\alpha} \eqno(2)$$
If these relations are applied to the observed H$\alpha$ fluxes 
for VB 10 and LP944-20, the upper limits for the reported 
non-detections in \ion{C}{4} and X-rays 
are near or even above the predicted values.  For example, 
\citet{l95} found an upper limit for VB 10 of 
$F_{C IV} < 6 \times 10^{-16}$
ergs cm$^{-2}$ s$^{-1}$ while the scaling relation gives a
predicted \ion{C}{4} line flux of 
$9 \times 10^{-16}$ ergs cm$^{-2}$ s$^{-1}$ based on the observed
H$\alpha$ line flux from \citet{h96}.  Recently, \citet{m02}
reobserved LP944-20 with XMM-Newton, again failing to detect
quiescent emission; however, this observation is only a factor of
3 more sensitive than the previous Chandra result of \citet{r00}.
Thus, only slight deviations from
the above relations would explain the current
non-detections, without requiring a major paradigm shift with 
regard to magnetic activity on late M dwarfs compared to earlier
M dwarfs.

The case for the absence of a quiescent transition region 
and corona in the lowest mass stars and brown dwarfs
thus rests on tenuous observational evidence.  The goal of
this investigation is to place stringent limits on the
presence of transition region emission using the sensitive
HST/STIS spectrograph with the FUV-MAMA detector. 
These observations provide
the first rigorous test of the presence of persistent quiescent
emission on very late type main sequence dwarfs
at temperatures well above chromospheric levels, and
by inference the ubiquitous existence of a hot outer
atmosphere.   

\section{Observations}

Our target list comprises the brightest active stars known at 
spectral types M7 (VB 8), M8 (VB 10) and M9 (LHS 2065).  These 
targets were chosen to give us the best possible detection limits 
for persistent quiescent emission.  The total exposure time on
each object corresponds to a \ion{C}{4} detection limit ten times fainter
than the expected \ion{C}{4} line strength scaled from earlier type
active M dwarfs (equation 1).

The Hubble Space Telescope was used with the STIS spectrograph and 
the FUV-MAMA detector to obtain 15 consecutive exposures of each target.
All observations were made using the 52$^{\prime\prime}$ long by
0.2$^{\prime\prime}$ wide slit and the G140L grating to give a spectral
resolution $R \equiv \lambda/\delta\lambda \sim 3600$ at 1500\AA .
For VB 8, spectral observations started at 12:38 UT on 2001 October 6.
Each G140L spectrum was integrated for approximately 5 minutes, and the total
exposure sequence spanned two orbits.  For LHS 2065, spectral observations
started at 5:23 UT on 2002 April 27.  Each G140L spectrum was integrated
for approximately 10 minutes, and the total exposure sequence spanned four
orbits.  The observing sequence for VB 10 was identical to that of LHS 2065
with G140L spectral observations starting at 5:03 UT on 2002 September 7.
Table 1 gives a detailed log of the STIS observations.

The data were downloaded from the HST archive, and
data reduction was performed with the CALSTIS package \citep{Lin99}
utilizing the STIS pipeline.  However, due to the relative faintness 
of our targets, the pipeline package was unable to properly locate the
spectrum of the star on the detector and extract it.  Therefore, we
performed the spectral extraction from the flatfielded science image
manually.  To accomplish this, we coadded all 15 exposures for each star
and examined the resulting image by eye, where it was easy to identify
bright emission lines such as \ion{C}{4} and \ion{Si}{4}.  We then
used the order trace from the STIS pipeline, applying an offset such 
that the
trace went through the emission lines from our object.  Examining each
of the 15 exposures individually, we verified that the order location
did not change substantially (more than a pixel) during any one 
exposure.  We then extracted the spectrum and background around this
offset order location using the same extraction width (11 pixels) used
by the pipeline to determine a count rate spectrum.  The count rate
spectrum was converted to a flux spectrum via the transformation
provided in the CALSTIS pipeline reduction.  We verified our procedure
by extracting spectra around the nominal order locations provided by the
CALSTIS pipeline for the 15 exposures of each source, recovering the same
(noise) spectrum produced by the STIS pipeline to within the (somewhat
large) photon statistics.

\section{Results}

The spectra extracted for each target from the sum of all 15 exposures
are shown in Figure 1.  
The emission lines we have identified are typical of those found in active
late-type stars \citep{a01}, and are very similar in relative strength to earlier 
type M dwarf transition region spectra, as shown by comparison to the
recent STIS/FUV-MAMA spectrum of the dM1e star AU Mic \citep{p00}.
The \ion{C}{4} doublet at 1550\AA\ is the strongest feature typically
observed from low mass stars in the wavelength range of study here and
is also the strongest line observed in these 3 stars.
For presentation purposes, we have applied Gaussian smoothing to the 
spectra with a FWHM of 3\AA .  This smoothing is not used in the flux 
measurements given
in Table 1.  The 1-$\sigma$ uncertainty limits are also indicated for
the smoothed spectra in Figure 1.  

Table 1 gives the observed \ion{C}{4} doublet line flux for each
exposure in all three stars.  The fluxes were determined by integrating
the line over 6.4\AA\ centered on the doublet; the integration limits
were set by visual examination of the individual spectra.
The significance of these detections can be estimated from
the off-source background in the images,
as discussed by \citet{k91}; see also \citet{b96}.
The image of each STIS
spectrum shows a clear variation in the background along the dispersion
direction, likely due to scattered light from the strong Ly-$\alpha$
airglow present.  The background shows no variation in the spatial
axis at the position of the \ion{C}{4} line.  To estimate the significance
of the \ion{C}{4} detection in each exposure, we step above and below
the position of the \ion{C}{4} line on the spectral image and extract the
number of background counts in a box the same size as that used to extract
the actual \ion{C}{4} line.  This produces approximately 100 estimates
of the background, to which we fit a Poisson distribution.  We then 
compare the net number of counts detected in the \ion{C}{4} line to this 
distribution in order to determine the significance of the detection.
In all but one exposure on LHS 2065, the \ion{C}{4} line is detected
above the 3$\sigma$ level.  For VB 8, the weakest detection is at the
3.1$\sigma$ level, the strongest at the 15.0$\sigma$ level, with a
mean significance of 7.7$\sigma$.  For VB 10, these three values
are 3.3$\sigma$, 314$\sigma$, and 39.1$\sigma$ respectively.  For
LHS 2065, the corresponding detections are 2.9$\sigma$, 17.6$\sigma$,
and 10.1$\sigma$.  The continuum adjacent to the \ion{C}{4} line 
is not detected in the same sized integration box at a significant level
in any exposure.  

Although these estimates show that the \ion{C}{4}
line is detected in all exposures, they do not represent the 
uncertainties in the measured flux.  Using the Bayesian
method described by \citet{k91}, we have estimated the
68.3\% confidence limits\footnote{The 68.3\% confidence limits
correspond to $\pm 1\sigma$ for a Gaussian distribution.  We have
not assumed a Gaussian distribution here, but these confidence
limits represent a typical benchmark.}.
The flux confidence limits for each spectrum are reported
in Table 1.

Time sequences of the \ion{C}{4} measurements, together with the
uncertainties, for each target are shown
in Figure 2.  There is no definite sign of long-lived flaring 
(\ion{C}{4} rise and decay) in the time sequences observed.
The second orbit on VB 10 shows a
strong flare at the end of the orbit (exposure 7), so the duration of
the event cannot be measured.  The \ion{C}{4} flux does appear to be
variable between exposures on all three stars.
Quiescent emission does not typically occur at a constant flux
level on active M dwarfs; similar variability
is commonly seen in the quiescent H$\alpha$ emission fluxes
\citep{pmsu3}.  This may be a sign of small scale flaring as a cause
of the quiescent emission.  Our \ion{C}{4} data suggest
a common origin for the chromospheric and transition region
emission, in conflict with the hypothesis that quiescent 
chromospheric emission is present but that transition region
emission is absent except during strong flares.

The \ion{C}{4} line strengths we measure are:
$1.4 \pm 0.1 \times 10^{-15} $ 
ergs s$^{-1}$ cm$^{-2}$ for VB 8;
$2.8 (1.4) \pm 0.1 (0.1) \times 10^{-15}$ ergs s$^{-1}$ cm$^{-2}$ for VB 10;
and $7.1 \pm 0.6 \times 10^{-16}$ ergs s$^{-1}$ cm$^{-2}$ for LHS 2065.
The values in parentheses for VB 10 represent the quiescent flux
and uncertainty measured from the total of all spectra excluding
the strong flare observed in exposure 7 (see Figure 2).
Note that when integrated over the entire observation for each target,
the \ion{C}{4} line is detected with sufficient counts that
uncertainties can be estimated assuming a Gaussian distribution 
for the source and the background \citep{b96}.

The scaling relation based on the H$\alpha$  emission in these
stars, which assumes that the \ion{C}{4} to H$\alpha$ ratio is the same
as for earlier type M dwarfs (equation 1), predicted 
values of 2 $\times 10^{-15}$ (VB 8),
9 $\times 10^{-16}$ (VB 10), and 1 $\times 10^{-15}$ (LHS 2065)
ergs s$^{-1}$ cm$^{-2}$.  These
are close to the observed values, with VB 8 and LHS 2065 having slightly
lower observed fluxes, while the observed VB 10 (quiescent) flux is
somewhat higher than predicted.
The VB 10 (quiescent) flux is also a factor of two higher than the upper
limit quoted in \citet{l95}.

\section{Summary}

Our conclusion is that
active, very low-mass, main-sequence stars (spectral
types M7-M9) can maintain a quiescent 
chromosphere and transition region that is similar 
to those observed in active, earlier type M dwarfs.  
Flaring may still be ultimately responsible
for the outer atmospheric heating in these stars, even
during ``quiescence'';  the role of flares (particularly micro-flares) 
in coronal and transition region heating is still not understood for
the Sun or for active stars \citep[for example]{k02,g03}.  
However, theories involving changes in atmospheric structure, 
dynamo heating and other mechanisms proposed
to explain the lack of quiescent activity in very low mass stars
and brown dwarfs \citep{f00,mb02}
must accommodate the persistent quiescent 
activity that we have observed at least through spectral type M9.

This research was made possible by grant HST-GO-9090
from the Space Telescope Science Institute, operated
by AURA for NASA.

\clearpage


\begin{figure}
\figurenum{1}
\epsscale{0.8}
\plotone{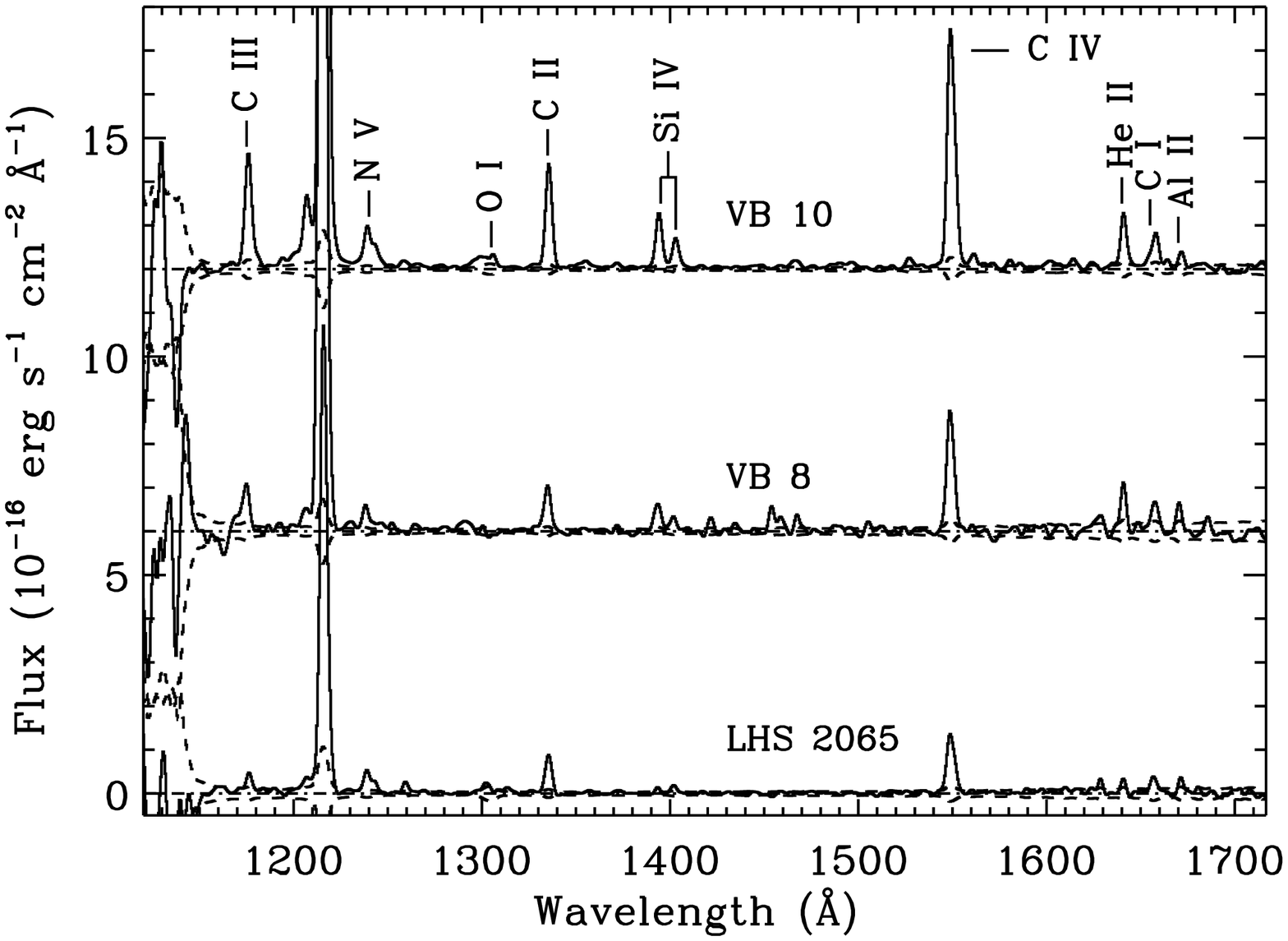}
\caption {The integrated HST STIS/MAMA-FUV 
spectra of all 3 targets are shown with the principal
emission lines identified.  VB 10 is offset by $12 \times 10^{-16}$
ergs s$^{-1}$ cm$^{-2}$ \AA$^{-1}$ and VB 8 is offset by
$6 \times 10^{-16}$ ergs s$^{-1}$ cm$^{-2}$ \AA$^{-1}$.  For each spectrum
the 1-$\sigma$ error bars are given by the dashed curves, while the zero 
flux line for each spectrum is shown by the dash-dot line.  These spectra
illustrate typical M dwarf transition region emission.}
\label{f4}
\end{figure}

\begin{figure}
\figurenum{2}
\epsscale{0.8}
\plotone{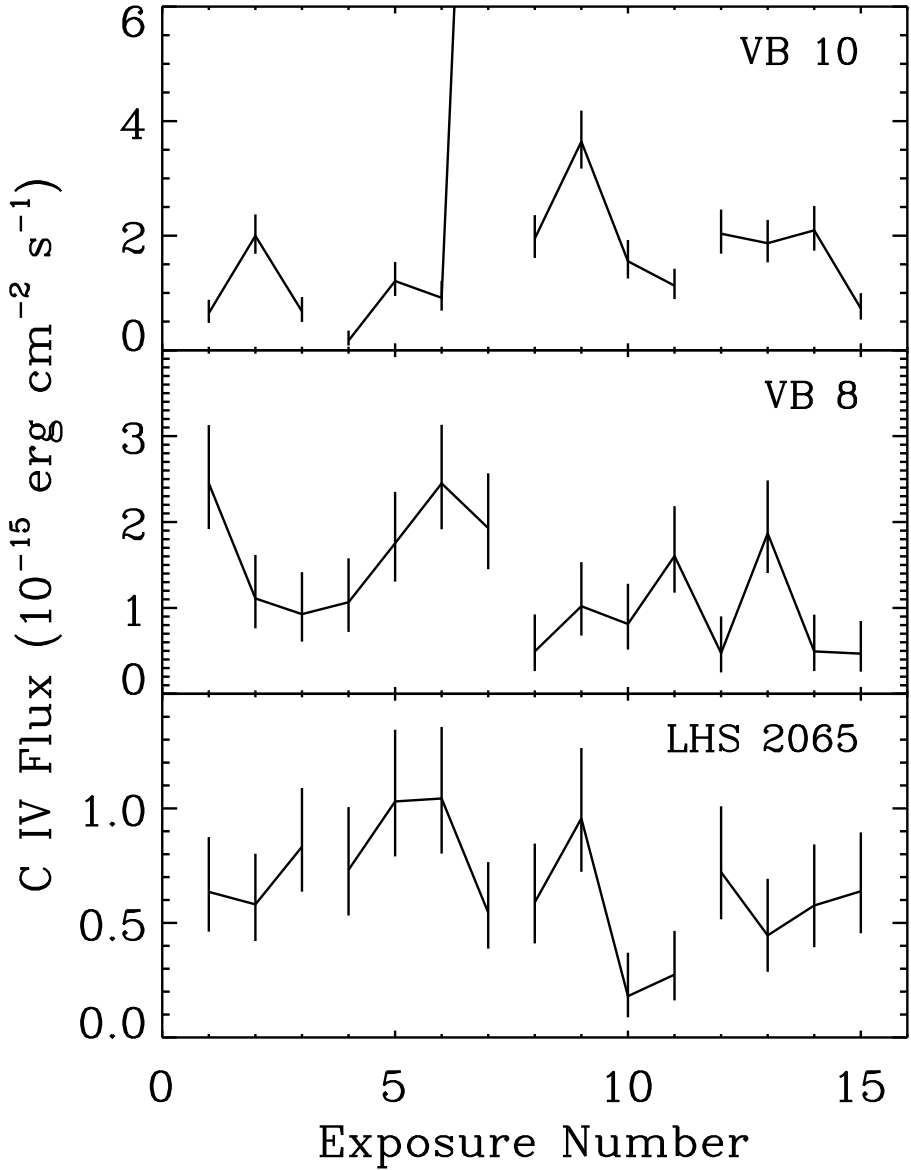}
\caption {The time series of \ion{C}{4} measurements for each star.
The break in the line connecting the different exposures corresponds
to the end of an HST orbit, producing a delay of $\sim 50$ minutes
before the next exposure.}
\label{f2}
\end{figure}

\clearpage

\begin{deluxetable}{cccccccccccc}
\tabletypesize{\scriptsize}
\tablewidth{17.5truecm}   
\tablecaption{Journal of Observations\label{observations}}
\tablehead{
   \colhead{ }&
   \multicolumn{3}{c}{VB 8}&
   \colhead{ }&
   \multicolumn{3}{c}{VB 10}&
   \colhead{ }&
   \multicolumn{3}{c}{LHS 2065}\\[0.2ex]
   \colhead{ }&
   \multicolumn{3}{c}{6 Oct. 2001}&
   \colhead{ }&
   \multicolumn{3}{c}{7 Sep. 2002}&
   \colhead{ }&
   \multicolumn{3}{c}{27 Apr. 2002}\\[0.2ex]
   \cline{2-4}
   \cline{6-8}
   \cline{10-12}
   \colhead{Exposure}&
   \colhead{Start}&
   \colhead{Exp.}&
   \colhead{$F_{\rm{C\ IV}} \times 10^{15}$}&
   \colhead{ }&
   \colhead{Start}&
   \colhead{Exp.}&
   \colhead{$F_{\rm{C\ IV}} \times 10^{15}$}&
   \colhead{ }&
   \colhead{Start}&
   \colhead{Exp.}&
   \colhead{$F_{\rm{C\ IV}} \times 10^{15}$}\\[0.2ex]
   \colhead{\#}&
   \colhead{(UT)}&
   \colhead{(sec)}&
   \colhead{(erg cm$^{-2}$ s$^{-1}$)}&
   \colhead{ }&
   \colhead{(UT)}&
   \colhead{(sec)}&
   \colhead{(erg cm$^{-2}$ s$^{-1}$)}&
   \colhead{ }&
   \colhead{(UT)}&
   \colhead{(sec)}&
   \colhead{(erg cm$^{-2}$ s$^{-1}$)}
}
\startdata
1&12:38:04&300&1.92-3.13 & &05:03:19&720&0.48-0.88 & &05:23:22&679&0.46-0.87\\
2&12:45:21&300&0.76-1.62 & &05:17:36&720&1.68-2.37 & &05:36:58&720&0.42-0.80\\
3&12:50:43&300&0.61-1.42 & &05:29:58&720&0.49-0.93 & &05:49:20&720&0.64-1.09\\
4&12:56:05&300&0.72-1.58 & &06:28:18&600&0.08-0.34 & &06:48:47&600&0.53-1.01\\
5&13:01:27&300&1.31-2.35 & &06:42:25&600&0.95-1.54 & &07:02:54&600&0.79-1.34\\
6&13:06:49&300&1.92-3.13 & &06:52:47&600&0.69-1.21 & &07:13:16&600&0.80-1.36\\
7&13:12:11&291&1.45-2.57 & &07:03:09&720&18.4-20.5 & &07:23:38&720&0.39-0.77\\
8&14:04:10&300&0.26-0.92 & &08:04:27&600&1.61-2.36 & &08:24:58&600&0.41-0.85\\
9&14:13:17&300&0.68-1.53 & &08:18:34&600&3.17-4.18 & &08:39:05&600&0.72-1.26\\
10&14:18:39&300&0.52-1.28 & &08:28:56&600&1.25-1.93 & &08:49:27&600&0.09-0.37\\
11&14:24:01&300&1.18-2.18 & &08:39:18&720&0.89-1.42 & &08:59:49&720&0.16-0.47\\
12&14:29:23&300&0.25-0.90 & &09:40:37&600&1.69-2.46 & &10:01:09&600&0.52-1.01\\
13&14:34:45&300&1.41-2.48 & &09:54:44&600&1.53-2.28 & &10:15:16&600&0.29-0.69\\
14&14:40:07&300&0.26-0.92 & &10:05:06&600&1.74-2.52 & &10:25:38&600&0.39-0.84\\
15&14:45:29&360&0.26-0.85 & &10:15:28&676&0.53-1.00 & &10:36:00&676&0.46-0.90\\
\enddata
\tablecomments{Values for the C IV flux give the 68.3\% confidence 
limits for the measured flux.}
\end{deluxetable}

\end{document}